# Electromagnetic form factors of nucleons in a relativistic three-quark model

M.A. Ivanov,* M.P. Locher, V.E. Lyubovitskij†

Paul Scherrer Institute
CH-5232 Villigen PSI

February 22, 1996

**Abstract**

We report the calculation of electromagnetic form factors of nucleons within a relativistic three-quark model with Gaussian shape for the nucleon-quark vertex. The allowed region for two adjustable parameters, the range parameter $\Lambda_N$ in the Gaussian and the constituent quark mass $m_q$, is obtained from fitting the data for magnetic moments and electromagnetic radii of nucleons. The shapes of electric and magnetic nucleon form factors calculated in this approach with the central parameters $m_q$=420 MeV and $\Lambda_N$=1.25 GeV are close to the data for momentum transfers $0 \leq Q^2 \leq 1$ GeV$^2$.

**PACS:** 12.35.Ht, 13.60.Fz

---

*Permanent address: Bogoliubov Laboratory of Theoretical Physics, Joint Institute for Nuclear Research, 141980 Dubna (Moscow region), Russia
†On leave of absence from: Bogoliubov Laboratory of Theoretical Physics, Joint Institute for Nuclear Research, 141980 Dubna (Moscow region), Russia. Permanent address: Department of Physics, Tomsk State University, 634050 Tomsk, Russia

# 1 Introduction

Electromagnetic properties of nucleons are an important source of information on the internal structure of baryons. The success of the nonrelativistic quark model for the description of static characteristics (masses, magnetic moments, etc.) and the results from deeply inelastic lepton scattering are a clear indication for the three-quark structure of nucleons.

In view of the difficulties of describing the nucleon as a relativistic three quark system rigorously, on the basis of QCD, many methods and models have been developed which implement important aspects of QCD at least partially. We mention a few. Some approaches [1],[2] are formulated using classical hadronic degrees of freedom (color singlets) without explicit reference to quarks and gluons. This holds in particular for chiral perturbation theory which addresses low energy scattering of hadrons [3]. In the Skyrmion model the baryon is considered to be a solitonic excitation of classical meson fields which are described by an effective chiral Lagrangian, see [4], e.g. The electromagnetic radii and magnetic moments of nucleons as well as the behavior of electromagnetic form factors up to 1 GeV$^2$ are described in this way.

In the context of quark models the nonrelativistic approximation is problematic even for the constituent-quark picture at low energies since the effective masses and the intrinsic momenta have the same order of magnitude. An attempt to implement relativistic invariance for the description of the electromagnetic properties of the nucleon is the covariant constituent quark model by Chung and Coester [5] which uses light front dynamics for the constituent quarks. In this framework the available data for $0 \leq Q^2 \leq 6$ GeV$^2$ has been well described with two adjustable parameters (constituent quark mass and confinement scale).

For large momentum transfers, $Q^2 >> m_N^2$, perturbative QCD predicts that the magnetic formfactor behaves as $G_M \sim 1/Q^4$ [6]. The use of QCD sum rules allows to extrapolate the electromagnetic form factors to low and moderate $Q^2$ by incorporating local quark-hadron duality [7]. Recall that the calculation of hadronic form factors by sum rules in their original version [8] is restricted to intermediate momentum transfers and does not work in the infrared region due to power corrections $1/Q^{2n}$. Therefore, in order to calculate magnetic moments of nucleons QCD sum rules in the external field approach have been introduced [9]. For a general method of calculating the nucleon magnetic form factors at small $Q^2$ see [10].

For applications at high energies the diquark model [11] has been proposed. In this model the proton is built from quarks and diquarks and the diquark is treated as a quasi-elementary particle. Fits to the data for $Q^2 > 3$ GeV$^2$ [12] with few parameters have been obtained.

A different line of attack uses both quark *and* hadronic degrees of freedom. Chiral potential models (see, for example, [13] and references therein) fall into this class. They use an effective confining potential for the quarks and a quark pion interaction which preserves chiral symmetry. Incorporating the lowest-order pionic correction, the magnetic moments of the nucleon octet have been calculated [13]. Alternative approaches [14]-[17] start from *effective Lagrangians* which describe the transition of hadrons to their constituent quarks, combined with some assumptions on the behavior of quarks at low energies and on the shape of the hadron-quark form factor. This approach is connected fairly directly to the difficult issues of hadronization and quark confinement in QCD.

In the present paper we extend the line of thought developed in [18] and [19] which use local hadron-quark or local hadron-quark-diquark vertices together with a confined quark propagator (no poles). It has been shown for these models that the electromagnetic form factors of nucleons tend to be somewhat above the data for $Q^2 < 1$ GeV$^2$. This might be a hint that the assumption of a *local* coupling of baryons with their constituents is not quite adequate. Introducing a *nonlocal* hadron-quark vertex may be a good way of effectively introducing other degreees of freedom (mesonic cloud, soft gluons, etc.). The Nambu-Jona-Lasinio models, NJL, with separable interactions are candidates for this purpose. Many successful applications of such models for low-energy pion physics exist (see, e.g., [14, 15] and references therein).

In [20, 21] the Lagrangian formulation of the NJL model with separable interaction has been given both for mesons and (for the first time) for baryons. The gauging of nonlocal interactions has been done by using the time-ordering P-exponent. The pion weak decay constant, the two-photon decay width, as well as the form factor of the $\gamma^*\pi^0 \to \gamma$ transition, the pion charge form factor, and the strong $\pi NN$ form factor have been calculated and good agreement with the data has been achieved with three parameters, the range parameters characterizing the size of mesons $\Lambda_M$ and baryons $\Lambda_B$, and the constituent quark mass $m_q$. In this paper we shall use the approach developed in [20, 21] in order to calculate the electromagnetic form factors of nucleons. First, we slightly modify the gauging of the nonlocal quark-hadron vertex by using a path-independent definition for the derivative of the time-ordering P-exponent [22, 23]. Second, we derive Feynman rules for the diagrams which describe the electromagnetic nucleon form factors. For simplicity, we use a Gaussian shape for the nucleon-quark vertex. The permissible range for the two adjustable parameters, the Gaussian range $\Lambda$ of the separable interaction and the constituent quark mass $m_q$, is obtained by fitting the experimental values of the magnetic moments and the electromagnetic radii. The momentum dependence of the electromagnetic form factors is then calculated and their shape is found to be close to the data, for the parameters obtained from fitting the static properties.

## 2  Model

We start with a brief review of our approach. We consider the hadron as being a bound state of relativistic constituent quarks with masses $m_q$ [20, 21]. The transition of hadrons into constituent quarks and *vice versa* is described by the corresponding interaction Lagrangian, and the momentum distribution of the constituents by an effective relativistic vertex function. Its shape is chosen to guarantee ultraviolet convergence. At the same time the vertex function is a phenomenological description of the long distance QCD interactions between quarks and gluons.

The interaction Lagrangian of nucleons with quarks is written as

$$\begin{aligned}
\mathcal{L}_N^{\text{int}}(x) &= \bar{N}(x) \int dy_1 \int dy_2 \int dy_3 \ \delta\left(x - \frac{y_1 + y_2 + y_3}{3}\right) F\left(\frac{1}{18}\sum_{i<j}(y_i - y_j)^2\right) \\
&\times \left\{g_N^V J_N^V(y_1, y_2, y_3) + g_N^T J_N^T(y_1, y_2, y_3)\right\} + \text{h.c.} \\
&= \bar{N}(x) \int d\xi_1 \int d\xi_2 F(\xi_1^2 + \xi_2^2) \\
&\times \left\{g_N^V J_N^V(x - 2\xi_1, x + \xi_1 - \sqrt{3}\xi_2, x + \xi_1 + \sqrt{3}\xi_2)\right. \\
&+ \left. g_N^T J_N^T(x - 2\xi_1, x + \xi_1 - \sqrt{3}\xi_2, x + \xi_1 + \sqrt{3}\xi_2)\right\} + \text{h.c.}
\end{aligned} \quad (1)$$

with $J_N^V$ and $J_N^T$ being the vector and tensor currents, respectively. The currents are symmetric under permutation of any two quarks

$$J_N^V(y_1, y_2, y_3) = \vec{\tau}\gamma^\mu\gamma^5 q^{a_1}(y_1)q^{a_2}(y_2)\tau_2\vec{\tau}C\gamma_\mu q^{a_3}(y_3)\varepsilon^{a_1 a_2 a_3}, \quad (2)$$

$$J_N^T(y_1, y_2, y_3) = \vec{\tau}\sigma^{\mu\nu}\gamma^5 q^{a_1}(y_1)q^{a_2}(y_2)\tau_2\vec{\tau}C\sigma_{\mu\nu} q^{a_3}(y_3)\varepsilon^{a_1 a_2 a_3} \quad (3)$$

The notation is as follows: $\{a\}$ stands for the color indices, $\sigma^{\mu\nu} = \frac{i}{2}[\gamma^\mu, \gamma^\nu]$, and $C = \gamma^0\gamma^2$ is the charge conjugation matrix, respectively.

The notation for the spin-flavour structure of the nucleon currents is the same as in [9, 24]. In particular, the proton quark currents in terms of isotopic $u$ and $d$ fields have the form

$$J_p^V = \gamma^\mu \gamma^5 d^a u^b C\gamma_\mu u^c \varepsilon^{abc}, \qquad (4)$$

$$J_p^T = \sigma^{\mu\nu}\gamma^5 d^a u^b C\sigma_{\mu\nu} u^c \varepsilon^{abc}. \qquad (5)$$

Both forms have been used separately in [18, 19] for studying the electromagnetic and strong properties of baryons. It was shown that the tensor current (5) is more suitable for the description of the data. For this reason we will use the tensor current (5) in the approach developed in this paper.

The vertex form factor $F(\xi_1^2 + \xi_2^2)$ characterizes the distribution of $u$ and $d$ quarks inside the nucleon. The choice of variables in the vertex function $F(\xi_1^2 + \xi_2^2)$ implies the use of the center of mass frame in which the coordinates of quarks are expressed as

$$y_1 = x - 2\xi_1 \quad y_2 = x + \xi_1 - \sqrt{3}\xi_2 \quad y_3 = x + \xi_1 + \sqrt{3}\xi_2,$$

$$\xi_1 = \frac{y_2 + y_3 - 2y_1}{6} \quad \xi_2 = \frac{y_3 - y_2}{2\sqrt{3}} \qquad (6)$$

with $\xi_1$ and $\xi_2$ being the Jacobi coordinates of the quarks.

The Fourier-transform of the vertex form factor is defined as

$$F(\xi_1^2 + \xi_2^2) = \int \frac{d^4k_1}{(2\pi)^4} \int \frac{d^4k_2}{(2\pi)^4} \exp(-ik_1\xi_1 - ik_2\xi_2) F(k_1^2 + k_2^2).$$

To introduce the electromagnetic field into the nonlocal nucleon-quark Lagrangian (1), we will use the time-ordering P-exponential [20, 21]. The interaction of quarks with the electromagnetic field is introduced by the standard minimal substitution.

Thus, we have the gauge invariant interaction Lagrangian

$$\begin{aligned}\mathcal{L}_N^{\text{int,GI}}(x) &= g_N \bar{N}(x) \int dy_1 \int dy_2 \int dy_3 \delta\left(x - \frac{y_1+y_2+y_3}{3}\right) F\left(\frac{1}{18}\sum_{i<j}(y_i-y_j)^2\right) \\ &\times \Gamma_1 \vec{\tau} \exp\left(ieQ\int_x^{y_1} dz_\mu A^\mu\right) q^{a_1}(y_1) \exp\left(ieQ\int_x^{y_2} dz_\mu A^\mu\right) q^{a_2}(y_2) \\ &\times C\Gamma_2 \tau_2 \vec{\tau} \exp\left(ieQ\int_x^{y_3} dz_\mu A^\mu\right) q^{a_3}(y_3)\varepsilon^{a_1 a_2 a_3} + \text{h.c.}\end{aligned} \qquad (7)$$

where $\Gamma_1$ and $\Gamma_2$ are spin matrices, and Q=diag(2/3,-1/3,-1/3) is the charge matrix of quarks.

This interaction Lagrangian generates nonlocal vertices which couple nucleons, photons and quarks. In general several diagrams contribute. For example, the electromagnetic form factors of nucleons are described by the standard triangle diagram (Fig.1a) and, additionally, by the bubble or contact diagrams (Fig.1b and Fig.1c).

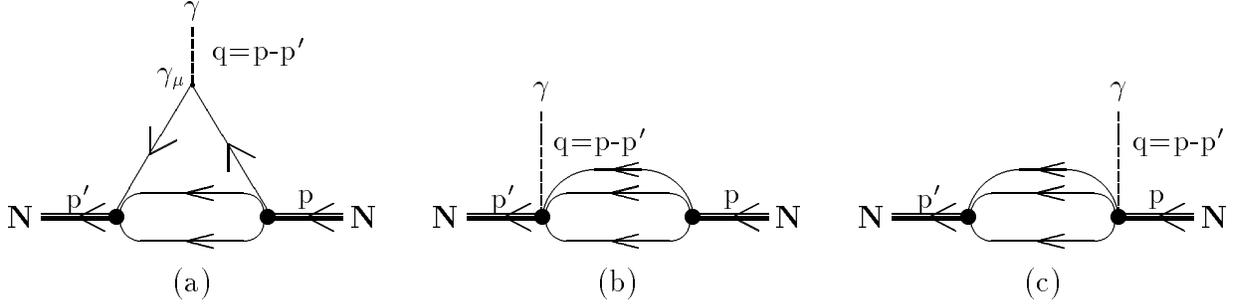

**Fig.1** Nucleon Form Factor. Triangle diagram (a), bubble diagrams (b) and (c).

At this stage the physical matrix elements describing the electromagnetic processes depend on the path which connects the lower and upper limits in the integral of the P-exponential. In the papers [20, 21] we used a straight line in our calculations. Here, we shall use a path-independent formalism which is based on the path-independent definition [22] of the derivative of the line integral. This formalism has been used in [23] for gauging nonlocal Lagrangians. We recall that the derivative of the line integral

$$I(x,y,P) \equiv \int_y^x dz_\mu A^\mu(z),$$

where $P$ is the path taken from $y$ to $x$, may be defined as [22]

$$\lim_{dx_\mu \to 0} dx_\mu \frac{\partial}{\partial x^\mu} I(x,y,P) = \lim_{dx_\mu \to 0} [I(x+dx,y,P') - I(x,y,P)]. \qquad (8)$$

Here $P'$ is obtained from the path $P$ by adding the extension $dx$ to the $x$ end.

Using the definition Eq.(8) leads to

$$\frac{\partial}{\partial x^\mu} I(x,y,p) = A_\mu(x). \qquad (9)$$

This means that the derivative of the $I(x,y,p)$ does not depend on the path used in the definition.

The method for deriving Feynman rules in the nonlocal chiral quark model with the electromagnetic P-exponential has been developed in [23]. The crucial point is to calculate the expression

$$F(-\partial_x^2)e^{-ipx}I(x,y,p)$$

using Eq.(9).

It is readily seen that

$$F(-\partial_x^2)e^{-ipx}I(x,y,P) = e^{-ipx}F(-\mathcal{D}_x^2)I(x,y,P)$$

where $\mathcal{D}_x^2 \equiv (\partial_x - ip)^2$. Thus we need to calculate

$$F(-\mathcal{D}_x^2)I(x,y,P) = \sum_{n=0}^{\infty}(-)^n\frac{F^{(n)}(0)}{n!}\mathcal{D}_x^{2n}I(x,y,P).$$

One finds that

$$\mathcal{D}_x^2 I(x,y,P) = (\partial_x A(x)) - 2i(pA(x)) - p^2 I(x,y,P) \equiv L(A) - p^2 I(x,y,P)$$

where $L(A) \equiv (\partial_x A(x)) - 2i(pA(x))$.

Then

$$(\mathcal{D}_x^2)^2 I(x,y,P) = (\mathcal{D}_x^2 - p^2)L(A) + (-p^2)^2 I(x,y,P)$$
$$(\mathcal{D}_x^2)^3 I(x,y,P) = ((\mathcal{D}_x^2)^2 - \mathcal{D}_x^2 p^2 + p^4)L(A) + (-p^2)^3 I(x,y,P)$$
$$\ldots$$
$$(\mathcal{D}_x^2)^n I(x,y,P) = \sum_{k=0}^{n-1}(\mathcal{D}_x^2)^{n-1-k}(-p^2)^k L(A) + (-p^2)^n I(x,y,P)$$
$$= n\int_0^1 dt[\mathcal{D}_x^2 t - p^2(1-t)]^{n-1}L(A) + (-p^2)^n I(x,y,P)$$

Finally, this leads to

$$F(-\mathcal{D}_x^2)I(x,y,P) = \int_0^1 dt F'[-\mathcal{D}_x^2 t + p^2(1-t)]L(A) + F(p^2)I(x,y,P)$$
$$= \int dq A_\mu(q)\left\{\left(q^\mu + 2p^\mu\right)e^{-iqx}\int_0^1 dt F'[(p+q)^2 t + p^2(1-t)] + F(p^2)\int_y^x dz^\mu e^{-iqz}\right\}.$$

We observe that the last term vanishes due to the delta function $\delta(x-y)$ in the Lagrangian.

For baryons, one has to calculate

$$F(-\partial_1^2 - \partial_2^2)e^{-ip_1\xi_1 - ip_2\xi_2} \int_y^{x+a_1\xi_1+a_2\xi_2} dz^\mu A_\mu(z).$$

By analogy we get

$$\int dq A_\mu(q) e^{-ip_1\xi_1 - ip_2\xi_2} \Big\{ e^{-iq(x+a_1\xi_1+a_2\xi_2)}\Big[(-ia_1)(a_1 q^\mu + 2p_1^\mu)F_1 + (-ia_2)(a_2 q^\mu + 2p_2^\mu)F_2\Big]$$
$$+F(p_1^2 + p_2^2) \int_y^{x+a_1\xi_1+a_2\xi_2} dz^\mu e^{-iqz} \Big\}.$$

with $F_1$ and $F_2$ being defined as

$$F_1 = \int_0^1 dt\, F'\Big[(p_2 + a_2 q)^2 + t(p_1 + a_1 q)^2 + (1-t)p_1^2\Big]$$
$$F_2 = \int_0^1 dt\, F'\Big[p_1^2 + t(p_2 + a_2 q)^2 + (1-t)p_2^2\Big]. \tag{10}$$

With this preparation the Feynman rules for the $\gamma NN$ vertex function may be derived. The details are given in Appendix A for the evaluation of the vertex in Fig.2 which connects a nucleon, a photon and three quarks, while appendix B gives the connection to the nucleon mass operator of Fig.3.

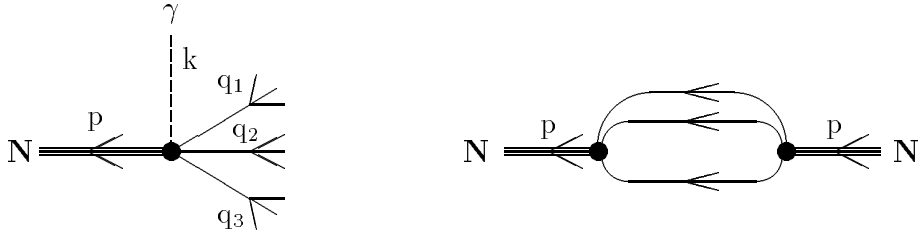

**Fig.2**. Nucleon photon three quark vertex.   **Fig.3**. Nucleon mass operator.

Now we turn to a discussion of the model aspects of our approach and the parameters involved. First, there is the vertex function in the Lagrangian (1). This is a free function except that it should make the Feynman diagrams ultraviolet finite, as we have mentioned before. In the papers [20, 21] we have found that the basic physical observables of pion low-energy physics depend only weakly on the choice of the vertex functions considered. Therefore, we choose in this paper a Gaussian vertex function for simplicity. In Minkowski space we write

$$F(k_1^2 + k_2^2) = \exp\left(\frac{k_1^2 + k_2^2}{\Lambda_N^2}\right)$$

where $\Lambda_N$ is the Gaussian range parameter which is related to the size of the nucleon.

The second input is the quark propagator. In the present paper we shall use the standard form, corresponding to a free fermionic field with mass $m_q$

$$< 0|T(q(x)\bar{q}(y))|0> = \int \frac{d^4k}{(2\pi)^4 i} e^{-ik(x-y)} S_q(k), \quad S_q(k) = \frac{1}{m_q - \slashed{k}} \tag{11}$$

Of course, the deficiency of such a choice is lack of confinement. This could be corrected by changing the analytic properties of the propagator. We leave that to a future study. For the time being we shall avoid the appearance of unphysical imaginary parts in the Feynman diagrams by restricting the calculations to $m_N < 3m_q$.

Thus, there are two adjustable parameters in our model: the constituent quark mass $m_q$ and the range parameter $\Lambda_N$. We shall determine them by fitting the electromagnetic properties of nucleons.

## 3  Nucleon Mass Operator and the $\gamma NN$ Vertex Function

Within the relativistic quark model described above we shall evaluate the magnetic moments, the charge radii and the behavior of the electric and magnetic form factors for $0 \leq Q^2 \leq 1$ GeV$^2$.

It is convenient to separate the contributions from the triangle diagram (Fig.1a) and the bubble diagrams (Fig.1b and 1c) as follows

$$\Lambda_\mu(p,p') = \frac{q_\mu}{q^2}[\Sigma_N(p) - \Sigma_N(p')] + \Lambda_{\mu,\Delta}^\perp(p,p') + \Lambda_{\mu,\text{bubble}}^\perp(p,p') \tag{12}$$

The functions $\Lambda_{\mu,\Delta}^\perp(p,p')$ and $\Lambda_{\mu,\text{bubble}}^\perp(p,p')$ are the modified triangle and bubble vertex functions which are orthogonal to the photon momentum $q^\mu \Lambda_\mu^\perp(p,p') = 0$. Thus the Ward-Takahashi identity following from (12) reads

$$q^\mu \Lambda_\mu(p,p') = \Sigma_N(p) - \Sigma_N(p'). \tag{13}$$

In particular, if $q = 0$ and both nucleons are on the mass shell, one finds

$$\Lambda_\mu(p,p) = \frac{\partial}{\partial p_\mu}\Sigma_N(p). \tag{14}$$

The expressions for the orthogonal parts of the vertex functions are

$$\begin{aligned}
\Lambda_{\mu,\Delta}^\perp(p,p') &= 216(g_N^T)^2 \int \frac{d^4k}{(2\pi)^4 i} \int \frac{d^4k'}{(2\pi)^4 i} \\
&\times F(12[k^2 + k'^2 + kk'])\, F(12[k^2 + k'^2 + kk'] + 4q^2 - 12kq) \\
&\times \Big[ 36Q Z_\mu^\perp(k,p,p') \Pi_{PP}(k,k',p) + 36Q\gamma_5 Z_\mu^\perp(k,p,p')\gamma_5 \Pi_{SS}(k,k',p) \\
&\quad + \vec{\tau}Q\vec{\tau}\sigma^{\alpha\beta}\gamma_5 Z_\mu^\perp(k,p,p')\sigma^{\rho\sigma}\gamma_5 \Pi_{TT}^{\alpha\beta,\rho\sigma}(k,k',p) \Big]
\end{aligned}$$

$$\begin{aligned}
\Lambda_{\mu,\text{bubble}}^\perp(p,p') &= 216(g_N^T)^2 \int \frac{d^4k}{(2\pi)^4 i} \int \frac{d^4k'}{(2\pi)^4 i} F(12[k^2 + kk' + k'^2]) k_\mu^\perp \\
&\times \frac{F(12[k^2 + k'^2 + kk'] + 4q^2 - 12kq) - F(12[k^2 + k'^2 + kk'])}{q^2 - 3kq} \\
&\times \Big[ 36Q S_q\!\left(k + \frac{p}{3}\right)\Pi_{PP}(k,k',p) + 36Q\gamma_5 S_q\!\left(k + \frac{p}{3}\right)\gamma_5 \Pi_{SS}(k,k',p) \\
&\quad + \vec{\tau}Q\vec{\tau}\sigma^{\alpha\beta}\gamma_5 S_q\!\left(k + \frac{p}{3}\right)\sigma^{\rho\sigma}\Pi_{TT}^{\alpha\beta,\rho\sigma}(k,k',p) \Big] \\
&+ (p \leftrightarrow p')
\end{aligned}$$

where $Z_\mu^\perp = S_q\!\left(k + p' - \frac{2p}{3}\right)\gamma_\mu^\perp S_q\!\left(k + \frac{p}{3}\right)$, $Q = \text{diag}(2/3,-1/3)$, $\vec{\tau}Q\vec{\tau} = \text{diag}(0,1)$,

$\gamma_\mu^\perp = \gamma_\mu - q_\mu \slashed{q}/q^2$, $k_\mu^\perp = k_\mu - q_\mu k\cdot q/q^2$ $(\gamma^\perp \cdot q = 0$ and $k^\perp \cdot q = 0)$.

$$\Pi_{PP}(k,k',p) = \text{tr}\left[\gamma_5 S_q\!\left(k' + \frac{p}{3}\right)\gamma_5 S_q\!\left(k + k' - \frac{p}{3}\right)\right],$$

$$\Pi_{SS}(k,k',p) = \text{tr}\left[S_q\!\left(k' + \frac{p}{3}\right) S_q\!\left(k + k' - \frac{p}{3}\right)\right],$$

$$\Pi_{TT}^{\alpha\beta,\rho\sigma}(k,k',p) = \text{tr}\left[\sigma^{\alpha\beta} S_q\!\left(k' + \frac{p}{3}\right)\sigma^{\rho\sigma} S_q\!\left(k + k' - \frac{p}{3}\right)\right]$$

The nucleon-three-quark coupling $g_N^T$ is calculated from the *compositeness condition* (see [16]) which means that the renormalization constant of the nucleon wave function is equal to zero, $Z_N = 1 - \Sigma'_N(m_N) = 0$, with $\Sigma_N$ being the nucleon mass operator

$$\Sigma_N(p) = 216(g_N^T)^2 \int \frac{d^4k}{(2\pi)^4 i} \int \frac{d^4k'}{(2\pi)^4 i} F^2\left(12[k^2 + k'^2 + kk']\right) \tag{15}$$

$$\times \left[12 S_q\left(k + \frac{p}{3}\right) \Pi_{PP}(k, k', p) + 12\gamma^5 S_q\left(k + \frac{p}{3}\right) \gamma^5 \Pi_{SS}(k, k', p)\right.$$

$$\left. + \sigma^{\alpha\beta}\gamma^5 S_q\left(k + \frac{p}{3}\right) \sigma^{\rho\sigma\beta}\gamma^5 \Pi_{TT}^{\alpha\beta,\rho\sigma}(k, k', p)\right]$$

The calculation of the electromagnetic vertex functions and of the nucleon mass operators is given in Appendix B.

## 4 Numerical Results and Discussion

We start with a summary of well-known definitions for the electromagnetic properties of the nucleon. The vertex functions in terms of Dirac $F_D^{(p,n)}$ and Pauli $F_P^{(p,n)}$ form factors are

$$M_\mu = \bar{N}^i(p') \Lambda_\mu^{ij}(p, p') N^j(p), \tag{16}$$

$$\Lambda_\mu^{ij}(p, p') = \gamma_\mu F_D^{ij}(q^2) - \frac{i}{2m_N}\sigma_{\mu\nu} q^\nu F_P^{ij}(q^2)$$

where $N(p)$ and $N(p')$ are the spinors corresponding to initial and final nucleon states with momentum $p$ and $p'$, respectively, and

$$\bar{N} = (\bar{p}\ \bar{n}), \qquad N = \begin{pmatrix} p \\ n \end{pmatrix} \tag{17}$$

$$F_D = \begin{pmatrix} F_D^p & 0 \\ 0 & F_D^n \end{pmatrix}, \qquad F_P = \begin{pmatrix} F_P^p & 0 \\ 0 & F_P^n \end{pmatrix}$$

The electric and magnetic (Sachs) form factors are obtained from

- Electric Form Factors $\qquad G_E^{(p,n)}(q^2) = F_P^{(p,n)}(q^2) + \dfrac{q^2}{4m_N^2} F_D^{(p,n)}(q^2)$

- Magnetic Form Factors $\qquad G_M^{(p,n)}(q^2) = F_P^{(p,n)}(q^2) + F_D^{(p,n)}(q^2)$

At zero recoil, $q^2 = 0$, these form factors satisfy the following normalization conditions

$$G_E^p(0) = 1, \quad G_E^n(0) = 0$$
$$G_M^p(0) = \mu_p, \quad G_M^n(0) = \mu_n$$

where $\mu_p$ and $\mu_n$ are the magnetic moments of the proton and the neutron, respectively.

The charge radii of the nucleons are given by

$$<r^2>_E^p = \frac{6}{G_E^p(0)} \left.\frac{dG_E^p(q^2)}{dq^2}\right|_{q^2=0} = \left[F_D^{p\prime}(0) + \frac{\mu_p - 1}{4m_N^2}\right] \tag{18}$$

$$<r^2>_E^n = 6\left.\frac{dG_E^n(q^2)}{dq^2}\right|_{q^2=0} = 6\left[F_D^{n\prime}(0) + \frac{\mu_n}{4m_N^2}\right] \tag{19}$$

$$<r^2>_M^N = \frac{6}{G_M^N(0)} \left.\frac{dG_M^N(q^2)}{dq^2}\right|_{q^2=0} = \frac{6}{\mu_N}\left[F_D^{N\prime}(0) + F_P^{N\prime}(0)\right] \tag{20}$$

We shall work in the limit of isospin invariance where the masses of $u$ and $d$ quarks are equal, $m_u = m_d = m_q$. In a first step we set the parameters of our model $m_q$ and $\Lambda_N$ to the well measured static properties of the nucleons, to the magnetic moments and to the charge radii [25, 26]. In a second step we shall explore the constraints from the shape of the nucleon electromagnetic form factors for the range $Q^2 \leq 1$ GeV$^2$.

In Table 1 the results for the static properties are given for various sets of parameters $m_q$ and $\Lambda_N$. As was mentioned above, the constituent mass $m_q$ is chosen to be more than one-third of the nucleon mass ($m_q > m_N/3$) to avoid an imaginary part in the Feynman diagrams. Its value is varied in the region from 315 MeV to 500 MeV. One of the important constraints is the ratio of magnetic moments of proton and neutron $R = |\mu_p/\mu_n|$ which is measured with high accuracy and equal to $R = 1.46$. Historically, the first success in describing this value has been achieved in a naive nonrelativistic $SU(6)$ quark model. The value $R_{SU(6)} = 1.5$ [27] is already close to experiment. It is seen from Table 1 that in our approach this value varies in the range $R=1.50$-1.57 when the quark mass is smaller than 400 MeV, and starts to be less than 1.5 for $m_q >$400 MeV. The electric neutron radius squared, $<r^2>_E^n$, is also sensitive to the choice of the quark mass. In this case as well the best agreement with experiment is achieved for $m_q >$400 MeV. As can be seen from (19) the neutron charge radius is a combination of the neutron Dirac form factor $F_D^{n\prime}(0)$ and the ratio $3\mu_n/2m_N^2$. Using the experimental value for the neutron magnetic moment the last term almost reproduces the experimental value for $<r^2>_E^n$. This means that the contribution coming from $F_D^{n\prime}(0)$ should be negligibly small ($F_D^{n\prime}(0) \simeq 0.008$). One can see from Table 2 that this value, disregarding its sign, is indeed small for $m_q >$ 400 MeV which confirms a consistent trend for the value of $m_q$. Therefore we consider $m_q =$ 400 MeV to be a lower bound for the constituent quark mass in our model. Our results for magnetic moments get worse for $m_q >$ 500 MeV (see, Table 1). The value of 500 MeV may therefore be considered an approximate upper bound for $m_q$.

Table 1. Static Nucleon Properties

| $m_q$, $\Lambda_N$ in GeV | $\mu_p$ | $\mu_n$ | $\left|\frac{\mu_p}{\mu_n}\right|$ | $r_E^p$ (fm) | $<r^2>_E^n$ (fm$^2$) | $r_M^p$ (fm) | $r_M^n$ (fm) |
|---|---|---|---|---|---|---|---|
| 0.315, 2.40 | 2.78 | -1.77 | 1.57 | 0.91 | -0.171 | 0.84 | 0.85 |
| 0.320, 2.25 | 2.79 | -1.79 | 1.56 | 0.88 | -0.169 | 0.80 | 0.81 |
| 0.325, 2.12 | 2.81 | -1.81 | 1.55 | 0.88 | -0.166 | 0.79 | 0.80 |
| 0.330, 2.00 | 2.82 | -1.82 | 1.55 | 0.88 | -0.162 | 0.79 | 0.80 |
| 0.335, 1.90 | 2.82 | -1.84 | 1.53 | 0.88 | -0.159 | 0.79 | 0.80 |
| 0.340, 1.80 | 2.82 | -1.85 | 1.52 | 0.89 | -0.155 | 0.80 | 0.80 |
| 0.350, 1.70 | 2.82 | -1.85 | 1.52 | 0.89 | -0.152 | 0.80 | 0.80 |
| 0.360, 1.60 | 2.83 | -1.86 | 1.52 | 0.89 | -0.148 | 0.81 | 0.81 |
| 0.370, 1.50 | 2.82 | -1.87 | 1.51 | 0.90 | -0.144 | 0.82 | 0.82 |
| 0.380, 1.40 | 2.82 | -1.87 | 1.51 | 0.91 | -0.139 | 0.83 | 0.83 |
| 0.390, 1.35 | 2.81 | -1.87 | 1.50 | 0.92 | -0.137 | 0.84 | 0.84 |
| 0.400, 1.30 | 2.81 | -1.87 | 1.50 | 0.92 | -0.135 | 0.84 | 0.84 |
| 0.410, 1.30 | 2.79 | -1.86 | 1.50 | 0.91 | -0.134 | 0.83 | 0.83 |
| 0.420, 1.25 | 2.79 | -1.86 | 1.50 | 0.92 | -0.132 | 0.84 | 0.84 |
| 0.430, 1.25 | 2.77 | -1.86 | 1.49 | 0.91 | -0.132 | 0.83 | 0.83 |
| 0.440, 1.25 | 2.76 | -1.85 | 1.49 | 0.90 | -0.131 | 0.82 | 0.82 |
| 0.450, 1.25 | 2.76 | -1.85 | 1.49 | 0.89 | -0.131 | 0.81 | 0.81 |
| 0.460, 1.20 | 2.75 | -1.85 | 1.49 | 0.90 | -0.129 | 0.82 | 0.82 |
| 0.470, 1.20 | 2.74 | -1.84 | 1.49 | 0.89 | -0.129 | 0.82 | 0.81 |
| 0.480, 1.15 | 2.74 | -1.84 | 1.49 | 0.90 | -0.127 | 0.83 | 0.82 |
| 0.490, 1.15 | 2.73 | -1.84 | 1.48 | 0.90 | -0.127 | 0.82 | 0.82 |
| 0.500, 1.15 | 2.73 | -1.84 | 1.48 | 0.89 | -0.127 | 0.82 | 0.81 |
| Exp. | 2.79 | -1.91 | 1.46 | 0.86±0.01 | -0.119±0.004 | 0.86±0.06 | 0.88±0.07 |

Table 2. Contribution of Dirac Form Factor to Nucleon Radii
comp. Eqs. (19)-(21)

| $m_q$, $\Lambda_N$ in GeV | $F_D^{p\prime}(0)$ in GeV$^{-2}$ | $F_D^{n\prime}(0)$ in GeV$^{-2}$ |
|---|---|---|
| 0.315, 2.40 | 3.04 | -0.229 |
| 0.320, 2.25 | 2.80 | -0.215 |
| 0.330, 2.00 | 2.80 | -0.176 |
| 0.340, 1.80 | 2.89 | -0.138 |
| 0.350, 1.70 | 2.87 | -0.125 |
| 0.360, 1.60 | 2.87 | -0.115 |
| 0.370, 1.50 | 2.95 | -0.085 |
| 0.380, 1.40 | 3.02 | -0.073 |
| 0.390, 1.35 | 3.10 | -0.055 |
| 0.400, 1.30 | 3.10 | -0.047 |
| 0.410, 1.30 | 3.04 | -0.045 |
| 0.420, 1.25 | 3.11 | -0.037 |
| Exp. (in GeV$^{-2}$) | 2.66 | 0.008 |

The electric radius of the proton and the magnetic radii of protons and neutrons depend fairly weekly on the parameters $m_q$ and $\Lambda_N$. The theoretical values for the neutron electric radius squared show the right trend as well, despite of the small size of this quantity. In Table 3 the best fit of the nucleon static properties for $m_q =$ 420 MeV and $\Lambda_N =$ 1.25 GeV is compared to other theoretical approaches such as QCD sum rules [9, 28], the nonlinear chiral model of [4], the relativistic invariant quark model in the light-front formalism of [29], the nonrelativistic quark model of [30], the relativistic quark model with chiral symmetry of [13]), the nonrelativistic quark model with QCD vacuum induced quark interactions of [31] and the MIT bag model [32].

Table 3. Static Properties Compared to Theoretical Approaches

| Approach | $\mu_p$ | $\mu_n$ | $r_E^p$, fm | $<r^2>_E^n$, fm$^2$ | $r_M^p$, fm | $r_M^n$, fm |
|---|---|---|---|---|---|---|
| Our | 2.79 | -1.86 | 0.92 | -0.132 | 0.84 | 0.84 |
| Ioffe [9] | 2.96 | -1.93 | | | | |
| Balitsky [28] | 2.80 | -1.95 | | | | |
| Meissner [4] | 2.77 | -1.84 | 0.97 | -0.25 | 0.94 | 0.94 |
| Aznauryan [29] | 2.811 | -1.848 | 0.811 | -0.094 | 0.825 | 0.781 |
| Isgur [30] | 2.79 | -1.86 | | | | |
| Baric [13] | 2.73 | -1.975 | 0.85 | | | |
| Dorokhov [31] | 2.20 | -1.47 | 0.831 | 0 | | |
| De Grand [32] | 1.90 | -1.27 | 0.728 | 0 | | |
| Exp. | 2.79 | -1.91 | 0.86±0.01 | -0.119±0.004 | 0.86±0.06 | 0.88±0.07 |

The shape of the electromagnetic nucleon form factors in the region $0 \leq Q^2 \leq 1$ GeV$^2$ for the same optimal values of the parameters, $m_q$=420 MeV and $\Lambda_N$=1.25 GeV, is shown in Figs.4-7. The normalized magnetic form factors of nucleons $G_M^N(Q^2)/G_M^N(0)$ are plotted in Fig.4 and Fig.5.

The short dashed line corresponds to the contribution of the triangle vertex function $\Lambda_{\mu,\Delta}^{\perp}(p,p')$ whereas the medium-long dashed line gives the contribution of the bubble or contact term $\Lambda_{\mu,bubble}^{\perp}(p,p')$. This contribution is roughly constant and not large. It matters however for the details of the fitting procedure. The total contribution is marked by the solid line. The long-dashed line is the dipole approximation $D(Q^2) = [1 + Q^2/0.71\text{GeV}^2]^{-2}$. It can be taken as a representation of the data which it fits quite well. One can see that our results are close to the dipole fit and hence to the data.

The electric form factors of the proton and the neutron are shown in Fig.6 and 7, respectively. The data in Fig.7 characterize the experimental situation for the neutron electric form factor. We remark that the behavior of the nucleon electromagnetic form factors is sensitive to the choice of the constituent quark mass, as we have checked. Increasing the constituent quark mass above 500 MeV and reducing the parameter $\Lambda_N$ below 1.15 GeV improves agreement with the static nucleon properties. However, the form factors at $Q^2$=1 GeV$^2$ are then definitely getting too low. Thus the value $m_q$=500 MeV is an upper limit in our model also from the point of view of the form factors.

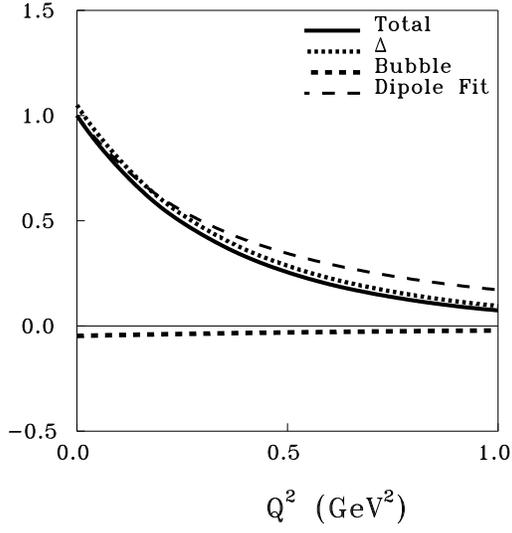
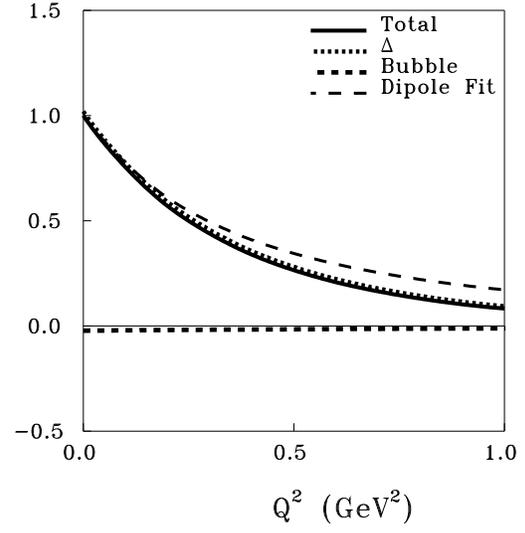

**Fig.4**. Magnetic form factor of proton.   **Fig.5**. Magnetic form factor of neutron.

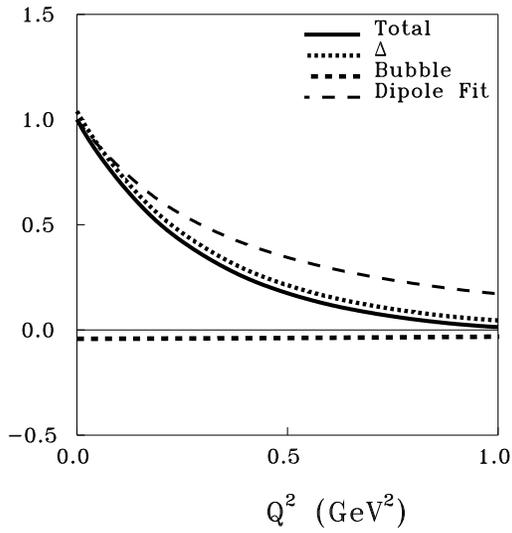
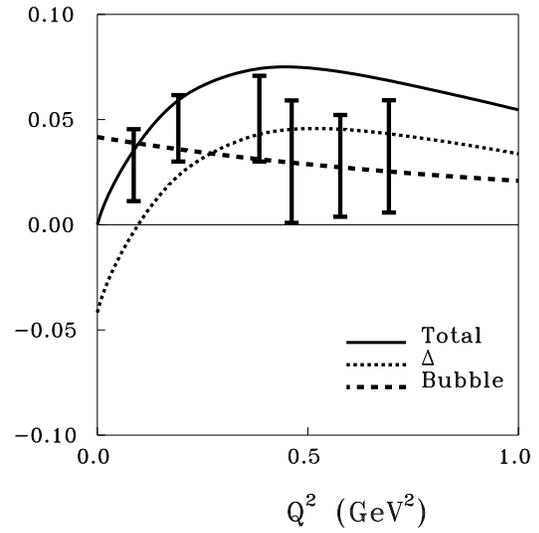

**Fig.6**. Electric form factor of proton.   **Fig.7**. Electric form factor of neutron.

## 5  Conclusion

The electromagnetic form factors of nucleons have been calculated within a relativistic three quark model with Gaussian shape for the nucleon-quark vertex, and standard (non-confined) quark propagators. Gauge invariance of the nonlocal hadron-quark interaction has been implemented by the path-independent definition for the derivative of the time-ordering P-exponent. The allowed region for the two adjustable parameters, the range parameter $\Lambda_N$ appearing in the Gaussian and the constituent quark mass $m_q$, has been obtained by fitting the data for the magnetic moments and the electromagnetic radii of the nucleons. The corresponding shapes of the electric and magnetic nucleon form factors for $m_q$=420 MeV and $\Lambda_N$=1.25 GeV are close to experiment for $0 \leq Q^2 \leq 1$ GeV$^2$. A large body of data is therefore described by only two free parameters. It remains to be seen whether a further improvement is possible by introducing confined propagators into the model.

## 6  Acknowledgements

We would like to thank P. Kroll, R. Rosenfelder, J. Bolz, H. Kamada and H.-Q. Zheng for useful discussions. This work was supported in part by the INTAS Grant 94-739 and by the Russian Fund of Fundamental Research 94-02-03463-a.

# 7  Appendix

## A. Feynman Rules for Nonlocal Electromagnetic Vertices

We illustrate how to obtain a nonlocal vertex which couples a nucleon, a photon and three quarks, see Fig.2. The analogous nonlocal vertices with mesons were introduced in [23]. Let $p$ denote the momentum of the outgoing baryon, $k$ the momentum of the outgoing photon, and $q_1, q_2, q_3$ the momenta of incoming quarks, respectively. We start from the gauge invariant Lagrangian describing the interaction of nucleons with quarks, Eq.(7). The vertex function $\Gamma^\mu(q_1 q_2 q_3 p k)$ corresponding to Fig.2 is

$$i(2\pi)^4 \delta^{(4)}(p - k - \sum_{i=1}^{3} q_i) \Gamma^\mu_{a_1 a_2 a_3}(q_1 q_2 q_3 p k) = \int dx_1 ... \int dx_5 e^{i(\sum_{i=1}^{3} q_i x_i - p x_4 - k x_5)} R^\mu_{a_1 a_2 a_3}(x_1...x_5) \quad (A.1)$$

where the function $R^\mu_{a_1 a_2 a_3}(x_1...x_5)$ is given by

$$R^\mu_{a_1 a_2 a_3}(x_1...x_5) = \left. \frac{\delta^5[\int dx \mathcal{L}^{\text{int,GI}}_N(x)]}{\delta A_\mu(x_5) \delta \bar{N}(x_4) \delta q^{a_1}(x_1)...\delta q^{a_3}(x_3)} \right|_{A_\mu = 0} \quad (A.2)$$

$$= e g_N \varepsilon^{a_1 a_2 a_3} (\Gamma_1) \times (C\Gamma_2) F\left[\frac{1}{18}\sum_{i<j}^{3}(x_i - x_j)^2\right] \delta\left(x_4 - \frac{1}{3}\sum_{i=1}^{3} x_i\right)$$

$$\times \left[(\vec{\tau} Q) \times (\tau_2 \vec{\tau}) \int_{x_4}^{x_1} dz^\mu + \vec{\tau} \times (Q \tau_2 \vec{\tau}) \int_{x_4}^{x_2} dz^\mu + \vec{\tau} \times (\tau_2 \vec{\tau} Q) \int_{x_4}^{x_3} dz^\mu\right] \delta(x_5 - z)$$

We define a typical integral as

$$I^\mu_{\text{typ}} = \int dx_1 ... \int dx_5 e^{i(\sum_{i=1}^{3} q_i x_i - p x_4 - k x_5)} F\left[\frac{1}{18}\sum_{i<j}^{3}(x_i - x_j)^2\right] \delta\left(x_4 - \frac{1}{3}\sum_{i=1}^{3} x_i\right) \int_{x_4}^{x_1} dz^\mu \delta(x_5 - z)$$

which may be written as

$$I^\mu_{\text{typ}} = 6\sqrt{3} \int dx \int d\xi_1 \int d\xi_2 F(\xi_1^2 + \xi_2^2) \exp\left[ix(\sum_{i=1}^{3} q_i - p)\right] \quad (A.3)$$

$$\times \exp\left[\xi_1(-2q_1 + q_2 + q_3) + i\xi_2\sqrt{3}(q_3 - q_2)\right] \int_{x}^{x - 2\xi_1} dz^\mu \exp[-ikz]$$

Using the identity

$$\int d\xi_1 \int d\xi_2 f(\xi_1,\xi_2) F(\xi_1^2 + \xi_2^2) = \int d\xi_1 \int d\xi_2 f(\xi_1,\xi_2) F(-\partial_{\xi_1}^2 - \partial_{\xi_2}^2) \delta(\xi_1)\delta(\xi_2)$$

$$= \int d\xi_1 \int d\xi_2 \delta(\xi_1)\delta(\xi_2) F(-\partial_{\xi_1}^2 - \partial_{\xi_2}^2) f(\xi_1,\xi_2)$$

one finds

$$I_{typ}^{\mu} = 6\sqrt{3} \int dx \int d\xi_1 \int d\xi_2 \; \delta(\xi_1)\delta(\xi_2) \exp\left[ix(\sum_{i=1}^{3} q_i - p)\right] F(-\partial_{\xi_1}^2 - \partial_{\xi_2}^2)$$

$$\times \left\{ \exp[i\xi_1(-2q_1 + q_2 + q_3) + i\xi_2\sqrt{3}(q_3 - q_2)] \int_{x}^{x-2\xi_1} dz^{\mu} \exp[-ikz] \right\}.$$

Finally, using the transformations described in Sec. 2 we obtain

$$I_{\text{typ}}^{\mu} = i(2\pi)^4 \delta^{(4)}(p - k - \sum_{i=1}^{3} q_i) 24\sqrt{3}(q_2 + q_3 - 2q_1 + k)^{\mu} M(q_1, q_2, q_3, k) \quad (A.4)$$

$$M(q_1, q_2, q_3, k) = \frac{F[\Delta(q_1, q_2, q_3, k)] - F[\Delta(q_1, q_2, q_3, 0)]}{\Delta(q_1, q_2, q_3, k) - \Delta(q_1, q_2, q_3, 0)}$$

$$\Delta(q_1, q_2, q_3, k) = 4[\sum_{i=1}^{3} q_i^2 + k^2 - \sum_{i<j}^{3} q_i q_j + k(q_2 + q_3 - 2q_1)].$$

Using the result for the typical integral (A.4) in Eq.(A.1) we can write the vertex function $\Gamma_{a_1 a_2 a_3}^{\mu}(q_1 q_2 q_3 pk)$ as

$$\Gamma_{a_1 a_2 a_3}^{\mu}(q_1 q_2 q_3 pk) = 24\sqrt{3} e g_N \varepsilon^{a_1 a_2 a_3} (\Gamma_1) \times (C\Gamma_2) \quad (A.5)$$

$$\times \left[ \vec{\tau}Q \times (\tau_2\vec{\tau})(q_2 + q_3 - 2q_1 + k)^{\mu} M(q_1, q_2, q_3, k) \right.$$

$$+ \vec{\tau} \times (Q\tau_2\vec{\tau})(q_3 + q_1 - 2q_2 + k)^{\mu} M(q_2, q_3, q_1, k)$$

$$\left. + \vec{\tau} \times (\tau_2\vec{\tau}Q)(q_1 + q_2 - 2q_3 + k)^{\mu} M(q_3, q_1, q_2, k) \right]$$

## B. The nucleon mass operator and the $\gamma NN$ vertex

To demonstrate the calculation of an electromagnetic vertex and nucleon mass operator we consider the two generic integrals

$$I_1(p_E^2, p_E'^2, Q^2) = \int \frac{d^4 k_E}{\pi^2} \int \frac{d^4 k_E'}{\pi^2} \prod_{i=1}^{4} \frac{1}{m_q^2 + P_{iE}^2} \qquad (B.1)$$

$$\times \quad F(-12[k_E^2 + k_E'^2 + k_E k_E']) F(-12[k_E^2 + k_E'^2 + k_E k_E'] - 4Q^2 + 12 k_E Q)$$

$$I_2(p_E^2, p_E'^2, Q^2) = \int \frac{d^4 k_E}{\pi^2} \int \frac{d^4 k_E'}{\pi^2} \int_0^1 dt \prod_{i=1}^{3} \frac{1}{m_q^2 + P_{iE}^2} F(-12[k_E^2 + k_E'^2 + k_E k_E'])$$

$$\times \quad F'(-12[k_E^2 + k_E'^2 + k_E k_E'] - t[4Q^2 - 12 k_E Q])$$

with the Euclidean momenta $P_{iE}$ being related to nucleon momenta $p_E, p_E'$ by

$$P_{1E} = k_E + p_E/3, \quad P_{2E} = k_E' + p_E/3, \quad P_{3E} = k_E + k_E' - p_E/3,$$

$$P_{4E} = k_E + p_E' - 2p_E/3, \quad Q = p_E - p_E'.$$

The first integral arises in the calculation of the triangle vertex function or the derivative of the nucleon mass operator (in this case $p_E = p_E'$) The second integral arises in the calculation of the bubble vertex functions.

Using the Feynman parametrization $\frac{1}{A} = \int_0^\infty d\alpha \exp(-\alpha A)$ and integrating over $k_E$ and $k_E'$ one finds for on mass-shell nucleons

$$I_1(-m_N^2, -m_N^2, Q^2) = \int_0^\infty ... \int_0^\infty \frac{d\alpha_1 ... d\alpha_4}{\Delta_1^2} \exp\left[-\frac{3Q^2}{\Delta_1}(2 + \alpha_{12})(1 + 2\alpha_{34})\right]$$

$$\times \quad \exp\left[-\frac{4m_N^2}{3\Delta_1}((1+\alpha_{34})(\alpha_1 - \alpha_2)^2 + (1+\alpha_1)(\alpha_{34} - \alpha_2)^2 + (1+\alpha_2)(\alpha_{34} - \alpha_1)^2)\right]$$

$$\times \quad \exp\left[-12\left(m_q^2 - \frac{m_N^2}{9}\right) \sum_{i=1}^{4} \alpha_i\right]$$

$$I_2(-m_N^2, -m_N^2, Q^2) = 12 \int_0^1 dt \int_0^\infty ... \int_0^\infty \frac{d\alpha_1 ... d\alpha_3}{\Delta_2^2} \exp\left[-\frac{3Q^2 t}{\Delta_2}(2 + \alpha_{12})(2 + 2\alpha_3 - t)\right]$$

$$\times \quad \exp\left[-\frac{4m_N^2}{3\Delta_2}((1+\alpha_3)(\alpha_1-\alpha_2)^2+(1+\alpha_1)(\alpha_3-\alpha_2)^2+(1+\alpha_2)(\alpha_3-\alpha_1)^2)\right]$$

$$\times \quad \exp\left[-12\left(m_q^2-\frac{m_N^2}{9}\right)\sum_{i=1}^{3}\alpha_i\right]$$

Introducing the notation $\alpha_{ij} = \alpha_i + \alpha_j$ we have

$$\Delta_1 = 3 + 2\sum_{i=1}^{4}\alpha_i + \alpha_{12}\alpha_{34} + \alpha_1\alpha_2, \quad \Delta_2 = 3 + 2\sum_{i=1}^{3}\alpha_i + \sum_{i<j}^{3}\alpha_i\alpha_j.$$

Making a change of variables

$$\alpha_1 = xs, \quad \alpha_2 = x(1-s), \quad \alpha_3 = yt, \quad \alpha_4 = y(1-t) \quad \text{in integral } I_1(-m_N^2, -m_N^2, Q^2)$$

$$\alpha_1 = xs, \quad \alpha_2 = x(1-s), \quad \alpha_3 = y, \quad \text{in integral } I_2(-m_N^2, -m_N^2, Q^2)$$

one finds

$$I_1(-m_N^2, -m_N^2, Q^2) = \int_0^\infty dx \int_0^\infty dy \int_0^1 ds \frac{xy}{\Delta^2} \exp\left[-\frac{3Q^2}{\Delta}(2+x)(1+2y)\right] \quad \text{(B.2)}$$

$$\times \quad \exp\left[-\frac{4m_N^2}{3\Delta}((1+y)(1-2s)^2x^2+(1+xs)(y-x(1-s))^2+(1+x(1-s))(y-xs)^2)\right]$$

$$\times \quad \exp\left[-12(x+y)\left(m_q^2-\frac{m_N^2}{9}\right)\right]$$

$$I_2(-m_N^2, -m_N^2, Q^2) = 12\int_0^\infty dx \int_0^\infty dy \int_0^1 ds \int_0^1 dt \frac{x}{\Delta^2} \exp\left[-\frac{3Q^2t}{\Delta}(2+x)(2+2y-t)\right] \quad \text{(B.3)}$$

$$\times \quad \exp\left[-\frac{4m_N^2}{3\Delta}((1+y)(1-2s)^2x^2+(1+xs)(y-x(1-s))^2+(1+x(1-s))(y-xs)^2)\right]$$

$$\times \quad \exp\left[-12(x+y)\left(m_q^2-\frac{m_N^2}{9}\right)\right]$$

where $\Delta = 3 + 2(x+y) + x^2s(1-s) + xy$.